\documentclass[conference]{IEEEtran}
\usepackage{cite}
\usepackage[utf8]{inputenc}
\usepackage[nolist]{acronym}
\usepackage{amsmath}
\usepackage{bm}
\usepackage{color}
\DeclareMathOperator{\Poisson}{Poisson}
\DeclareMathOperator{\R}{R}
\DeclareMathOperator{\D}{D}
\DeclareMathOperator{\M}{M}
\usepackage{textcomp}
\usepackage{siunitx}
\DeclareSIUnit\pixel{pixel}
\DeclareSIUnit\voxel{voxel}
\DeclareSIUnit\photon{photons}
\DeclareSIUnit\iterations{iterations}
\newcommand{\SIr}[3]{\SI[round-mode=places, round-precision=#1]{#2}{#3}}
\newcommand{\numr}[2]{\num[round-mode=places, round-precision=#1]{#2}}
\DeclareUnicodeCharacter{00B5}{\textmu}
\DeclareUnicodeCharacter{2212}{\ensuremath{-}}
\DeclareUnicodeCharacter{00B0}{\SIUnitSymbolDegree}
\DeclareUnicodeCharacter{22C5}{\ensuremath{\cdot}}
\usepackage{pgf}
\usepackage{contour}
\contourlength{0.5pt}
\usepackage{tabu}
\usepackage{fully3dvars}
\usepackage{fdkvars}
\newcommand{\cs}{$y =$ \SI{0}{\milli\meter}}
\newcommand{\as}{$y =$ \SI{60}{\milli\meter}}

\begin{document}

\title{High-Fidelity Modeling of Shift-Variant Focal-Spot Blur for High-Resolution CT}

\author{Steven~Tilley~II,
Wojciech~Zbijewski, J.~Webster~Stayman}

\begin{acronym}
	\acro{CBCT}{Cone-Beam CT}
	\acro{MDCT}{Multi-Detector CT}
	\acro{MBIR}{Model-Based Iterative Reconstruction}
	\acro{FOV}{Field of View}
	\acro{SAD}{Source-Axis Distance}
	\acro{SDD}{Source-Detector Distance}
	\acro{GPL}{Gaussian Penalized-Likelihood}
	\acro{GPL-I}{GPL with Identity blur}
	\acro{GPL-SI}{GPL with Shift-Invariant blur}
	\acro{GPL-SV}{GPL with Shift-Variant blur}
	\acro{RMSE}{Root Mean Squared Error}
	\acro{ROI}{Region Of Interest}
\end{acronym}

\maketitle

\begin{abstract} 	Dedicated application-specific CT systems are popular solutions to high-resolution clinical needs.
	Some applications, such as mammography and extremities imaging, require spatial resolution beyond current capabilities.
	Thorough understanding of system properties may help tailor system design, acquisition protocols, and reconstruction algorithms to improve image quality.
	As resolution requirements increase, more accurate models of system properties are needed.
	Using a high-fidelity measurement model,
	we analyze the effects of shift-variant focal spot blur due to depth-dependence and anode angulation on image quality throughout the three-dimensional field of view of a simulated extremities scanner.
	A model of the shift-variant blur associated with this device is then incorporated into a Model-Based Iterative Reconstruction (MBIR) algorithm, which is then compared to FDK and MBIR with simpler blur models (i.e., no projection blur and shift-invariant projection blur) at select locations throughout the field of view.
	We show that shift-variant focal spot blur leads to location-dependent imaging performance.
    Furthermore, changing the orientation of the X-ray tube alters this spatial dependence.
	The analysis suggests methods to improve imaging performance based on specific image quality needs.
	For example, for small region of interest imaging, a transaxial X-ray tube orientation provides the best local image quality at a specific location, while for large objects an axial X-ray tube orientation provides better image quality uniformity. 
    The results also demonstrate that image quality can be improved by combining accurate blur modeling with MBIR.
	Specifically, across the entire field of view, MBIR with shift-variant blur modeling yielded the best image quality, followed by MBIR with a shift-invariant blur model, MBIR with an identity blur model, and FDK, respectively. These results suggest a number of opportunities for the optimization of imaging system performance in the hardware setup, the imaging protocol, and the reconstruction approach. While the high-fidelity models used here are applied using the specifications of a dedicated extremities imaging system, the methods are general and may be applied to optimize imaging performance in any CT system.
	\looseness=-1
	
\end{abstract}

\section{Introduction}
    There are a number of medical applications that demand high spatial resolution for effective diagnosis including quantitative imaging of bone health, detection of calcifications in mammography, and temporal bone imaging for otolaryngology.
	A number of dedicated imaging systems have been designed with specific high-resolution applications in mind. 
	For example, dedicated \ac{CBCT} systems have been designed for mammography \cite{Lai2007, Kwan2007}, extremities imaging \cite{Carrino2014, marinetto_quantification_2016}, and head imaging \cite{xu_technical_2016}.
	Such devices have highly variable system designs based on their application, and often choose different tradeoffs based on the high-resolution needs, necessities for compact designs, cost, etc.
	In particular, many \ac{CBCT} devices are based on flat-panel detectors which are found in a wide range of pixel sizes, coverage areas, and imaging performance.
	X-ray tubes in such systems are also highly variable --- with engineers often choosing compact fixed anode designs over larger rotating anode sources and using very wide cone angles to cover wide area detectors.
	With all of these flexibilities, \ac{CBCT} imaging properties exhibit strong system-dependence.
	\looseness=-1
    
	Understanding and modeling imaging properties is particularly important when optimizing a high resolution design.
	Spatial resolution requirements can drive selection of particular hardware elements, system geometries, acquisition protocols, and reconstruction algorithms.
	Traditional analysis that has been appropriate for conventional \ac{MDCT} (e.g., focusing on single, central slice image quality criteria at the center of the imaging \ac{FOV}) may not be appropriate or optimal for \ac{CBCT}. 
	Improved imaging system models aid in understanding limitations and improving performance. 
	In this work we focus on the modeling and analysis of shift-variant imaging properties of high-resolution CT systems.
	In particular, we implement a general model for an extended X-ray focal spot to investigate imaging properties throughout a large \ac{FOV} system.
	These models are applied to a prototype extremities \ac{CBCT} system design and investigated with different variations in the imaging chain (e.g., conventional versus advanced reconstruction methods).
	\looseness=-1

	The X-ray source can be a significant source of location-dependent image quality in \ac{CBCT}.
	X-ray tubes emit X-rays from a small area (focal spot) on a tungsten anode.
	To enable better heat dissipation (and permit higher current settings for lowering noise or decreasing acquisition time), larger focal spots are often employed.
	However, the anode is angled such that the focal spot has a small cross-sectional area when viewed from isocenter.
	Due to this angulation, the apparent size and shape of the focal spot can vary dramatically with location, contributing different amounts of blur to data depending on location.
	The location-dependence is more pronounced for larger cone- and fan-angles.
	Additionally, blur induced by the focal spot is subject to variable magnification for positions parallel to the source-detector axis.
	Thus, source blur due to the X-ray focal spot is complex with significant potential shift-variance throughout the \ac{FOV}.
	\looseness=-1

	The reconstruction algorithm is an important part of the CT imaging chain and can also have a dramatic effect on the imaging properties of a device.
	So-called \ac{MBIR} algorithms, which minimize an objective function based on system and noise models, have demonstrated improved imaging performance over traditional direct reconstruction approaches.
	\ac{MBIR} methods also provide a natural means to incorporate advanced measurement models (e.g., focal spot effects) directly in the reconstruction algorithm, resulting in simultaneous deblurring and image reconstruction.
    Thus, in addition to aiding system and acquisition protocol design, more accurate \ac{CBCT} forward models may improve reconstruction algorithm performance.
	\looseness=-1

	Shift-variant blur models have been incorporated in nuclear imaging \ac{MBIR} \cite{Tsui1987, mumcuoglu_accurate_1996} and in fan-beam \ac{MDCT} model-based sinogram restoration \cite{LaRiviere2007}.
    In \cite{tilley:16:msv, tilley_ii_penalized-likelihood_2017}, a general \ac{MBIR} framework that accounts for both source and detector blur as well as spatial noise correlations in projection measurements was introduced.
	In \cite{tilley:16:msv} a simple rectangular shift-variant focal spot blur model was incorporated into this algorithm, which was
	applied to simulated phantom data and to physical \ac{CBCT} test bench measurements. 
	While that study showed significant image quality improvements when \ac{MBIR} incorporates a space-variant blur model, that preliminary study did not include a depth-dependent model of focal spot blur and was limited to an investigation of a \ac{CBCT} system with a transaxially oriented X-ray tube (i.e., with the anode-cathode axis placed parallel to the plane of rotation) and reconstructions at only a few locations in the \ac{FOV}.  
\looseness=-1

	In this work, we analyze reconstruction image quality throughout the three-dimensional \ac{FOV} of a system with shift-variant focal spot blur and shift-invariant scintillator blur.
	A small trabecular bone phantom was used to ``probe'' image quality at locations throughout the \ac{FOV} --- data generated using a high-fidelity forward model was reconstructed with various reconstruction algorithms.
	The bone phantom was reconstructed at many locations throughout the \ac{FOV} to compare relative imaging performance.
	``Sweet spots'' within the \ac{FOV} where the finest spatial resolution details are discernible are identified, as are problem spots where focal spot effects contribute to lower resolution.
	FDK and both traditional and advanced \ac{MBIR} approaches are applied to see how imaging performance can be improved (and by how much) at different locations.
	Two X-ray tube orientations were considered: the anode-cathode axis oriented parallel to the axis of rotation (axial) and the anode-cathode axis oriented perpendicular to the axis of rotation (transaxial).
    The results of this study illustrate some of the spatial variations in imaging performance as well as potential reconstruction methods for improvement.
	The different forms of location-dependent image quality may suggest different design strategies for different applications depending on where the highest spatial resolution is required within the \ac{FOV}.
\looseness=-1

\section{Methods}

\begin{table}
	\centering
	\caption{Description of variables and symbols}\label{tab:vars}
	\begin{tabu}{c X c}
		Symbol & Description & Value \\
		\hline
		$\boldsymbol{\mu}$ & Vector of attenuation values & --- \\
		$\mathbf{y}$ & Vector of measurement data & --- \\
		$\mathbf{K}_Y$ & Measurement covariance matrix & --- \\
		$\mathbf{B}_d$ & Scintillator blur matrix & --- \\
		$\mathbf{B}_s$ & Focal-spot blur matrix & --- \\
		$\smash{\mathbf{A}_k}$ & System matrix with focal spot centered at sourcelet $k$ & --- \\
        $\smash{\mathbf{A}}$ & Reconstruction system matrix without sourcelets & --- \\
		$w_k$ & Relative intensity of sourcelet $k$ & --- \\
        $I_0$ & Barebeam source intensity & \SI{\numphotons{}}{\photon} \\
		$\M$ & Data binning operator (average over sub-pixels) & --- \\
        $H$ & Scintillator blur model parameter & \SIr{3}{\detH{}}{\square\milli\meter} \\
		$a$ & Scintillator blur Gaussian blur fraction & \numr{3}{\detgauss{}} \\
		$q$ & Radial frequency argument in blur model & --- \\
		$\sigma$ & Gaussian blur width parameter & \SIr{3}{\detsigma{}}{\per\milli\meter} \\
		$\boldsymbol{\sigma}_{ro}$ & Photon-equivalent readout noise & \SIr{3}{\readoutnoisesigma{}}{\photon} \\
		$\alpha$ & Anode angle & \ang{\anodeangle{}} \\
		SDD & Source-detector distance & \SI{\SDD{}}{\milli\meter} \\
		SAD & Source-axis distance & \SI{\SAD{}}{\milli\meter} \\
		$\R$ & Regularization function & --- \\
		$\D\{\cdot\} $ & Operator that places vector argument in a diagonal matrix & --- \\
	\end{tabu}
\end{table}

We would like to characterize high-resolution CT system performance with highly accurate forward models both in the simulation of data and incorporated into advanced reconstruction methods.
Performing this characterization over the entire \ac{FOV} is a computational challenge due to the small pixel sizes, small voxel sizes, focal spot complexities (shift-variance, depth-dependence), etc.
Thus, a strategy for local investigation within the larger \ac{FOV} was devised.
This strategy is discussed in the following subsections.

\subsection{Phantom and Data Generation}

\begin{figure}
	\centering
	\input{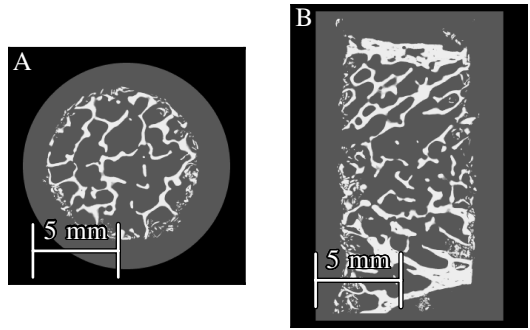}
	\caption{Axial (A) and Coronal (B) slice of trabecular bone phantom. The phantom consists of bone (\SIr{3}{\boneatten{}}{\per\milli\meter}) surrounded by fat (\SIr{3}{\fatatten{}}{\per\milli\meter}).}
	\label{fig:phantom}
\end{figure}

	A µCT scan of a human iliac crest biopsy sample was thresholded and used to create a realistic and clinically pertinent digital phantom (Fig.~\ref{fig:phantom}).
	The volume was binned to \SIr{3}{\phantomdsfivexfivespacing{}}{\milli\meter} cubic voxels prior to propagation through the imaging chain model.
	The digital phantom is intentionally small, serving as a high-resolution ``probe'' that can be scanned in various positions throughout the entire \ac{FOV}.
	The small support of the digital phantom permits both computationally efficient data generation and reconstruction for a detailed analysis of regional performance.
	\looseness=-1
    
\begin{figure}
	\centering
	\input{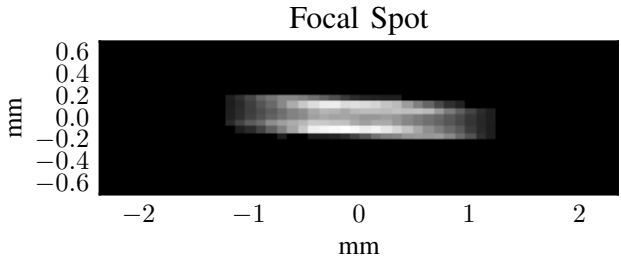}
	\caption{Focal spot model for data generation.
	         Each pixel represents a sourcelet, and the relative intensity indicates the relative weight of that sourcelet's measurements.
			 Due to the anode angle, less sampling was required along the long axis of the focal spot, resulting in anisotropic sourcelets.}
	\label{fig:focalspot}
\end{figure}

	For a realistic system characterization, a pinhole image of an X-ray focal spot from an IMD RTM 37 source (IMD, Grassobbio, Italy) was used to form a two-dimensional focal spot model for simulation.
	This X-ray tube has a $17.5^{\circ}$ anode angle and a nominal focal spot of \SI{0.6}{\milli\meter}.
    The source distribution on the anode is shown in Fig.~\ref{fig:focalspot}.
	This model was used to generate the shift-variant apparent focal spot distributions based on position within the \ac{FOV}.
	Sampling of the source distribution into ``sourcelets'' for projection and scaling by regional focal spot intensity was anisotropic (\SIr{3}{\fsx{}x\fsy{}}{\milli\meter}) with finer sampling along the short axis of the source and coarser sampling on the long axis since oblique views of the source result in finer sampling of the apparent focal spot.
	\looseness=-1

	Measurement data were generated according the following forward model:
	\begin{equation}
	\mathbf{y} = \mathbf{B}_d \Poisson \left( I_0 \M \left[ \sum_k w_k \exp \left( -\mathbf{A}_k
	\boldsymbol{\mu} \right) \right] \right) +
	\mathcal{N} \left( \mathbf{0}, \boldsymbol{\sigma}_{ro} \right). \label{eq:datagen}
	\end{equation}
	A description of all variables, symbols, and nominal values can be found in Table~\ref{tab:vars}.
	In short, at the core of (\ref{eq:datagen}), a volume of attenuation values, $\boldsymbol{\mu}$, is forward projected (on a fine grid of sub-pixels) for each focal spot sourcelet, $k$.
	The results are subsequently summed to form pre-detection projections.
	These finely sampled projections are binned using the operator $\M$ to account for non-linear partial-volume effects.
	In this work, data were binned by a factor of two in each direction, for a detector pixel pitch of \SI{\pixelpitch{}}{\milli\meter}.
	Pre-detection, incoming photons are presumed to be Poisson-distributed and undergo a detector blur based on the operator $\mathbf{B}_d$.
	This blur operation presumes a scintillator blur whose MTF is modeled by a Gaussian-Lorentzian mixture given by
	\begin{equation}
		H = a \exp(-q^2 / \sigma^2) + (1 - a)(1 + H q^2)^{-1}.
		\label{eq:detmtf}
	\end{equation}
	Lastly, measurements are subject to an additive uniform Gaussian readout noise. 
	\looseness=-1

	\subsection{Reconstruction}

\begin{figure*}[t!]
	\centering
	\input{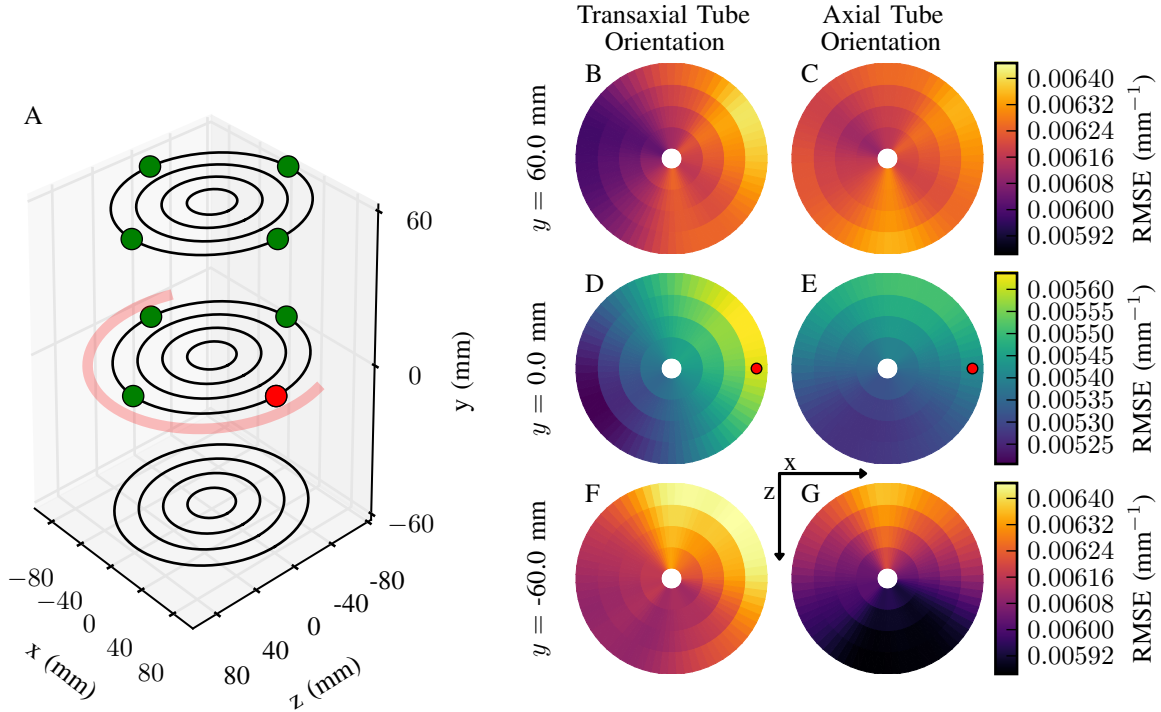}
	\caption{A: Schematic of \ac{FOV}.
			 Each ring was sampled using the FDK algorithm.
			 For a given sample point (for example the red dot), the phantom was placed at that location and scanned using a short scan with the source centered at ($0$, $0$, $\SAD{}$) and the detector centered at ($0$, $0$, -$\SDDmSAD{}$).
			 The angular range covered by the source is represented by the red arc.
			 The data were then reconstructed using FDK.
			 B: \ac{RMSE} of FDK reconstructions for each sample point (red dots indicate corresponding location in A).
			 }
	\label{fig:fdk}
\end{figure*}

	While various reconstruction methods are explored, in all cases, data were reconstructed with \SIr{4}{\voxelspacing{}}{\milli\meter} voxels.
	The phantom was placed \SIlist{20;40;60;80}{\milli\meter} from the axis of rotation and offset from the positive $x$ axis \ang{0} to \ang{355} in \ang{5} increments (rotated about the $y$ axis, see Fig.~\ref{fig:fdk}a).
	At the center of the short scan, the focal spot was at $z =$ \SI{\SAD{}}{\milli\meter} and the detector was at $z =$ \SI{-\SDDmSAD{}}{\milli\meter}.
	The source and detector rotated about the $y$ axis during scans.
	We conducted this experiment at three planes: the central plane (\cs{}) and \SI{60}{\milli\meter} above and below the central plane ($y = \pm 60$ mm).
	FDK reconstructions used a ramp filter with no apodization and a cutoff at the Nyquist frequency.
	Additionally, the X-ray tube's anode-cathode axis was modeled in two different orientations: {\em{Transaxial}} with the anode-cathode axis parallel to the plane of rotation; and {\em{Axial}} with the anode-cathode axis parallel to the axis of rotation.
	\looseness=-1

	We also reconstructed the phantom with three \ac{GPL} \ac{MBIR} methods: \ac{GPL-I}, \ac{GPL-SI}, and \ac{GPL-SV}.
	Each method minimized the following objective function
	\begin{equation}
		\psi = \left \| \mathbf{y} - \mathbf{B}_d\mathbf{B}_s I_0 \exp(-\mathbf{A} \boldsymbol{\mu}) \right \|_{\mathbf{K}^{-1}_Y} + \beta \R(\boldsymbol{\mu}) \label{eq:objective}
							\end{equation}
	where the covariance of measurement data is given by
	\begin{equation}
		\mathbf{K}_Y = \mathbf{B}_d \D\{\mathbf{y}\} \mathbf{B}_d^T + \D\{\boldsymbol{\sigma}_{ro}\}
		\label{eq:covariance}
	\end{equation}
	This reconstruction objective includes blur models for both the source ($\mathbf{B}_s$) and detector scintillator ($\mathbf{B}_d$) as well as a noise model with spatially correlated measurements (due to scintillator blur).
	A standard reconstruction system matrix ($\mathbf{A}$, without sourcelets) is used.
	The general \ac{GPL} algorithm to solve this objective was presented previously in \cite{tilley:16:msv, tilley_ii_penalized-likelihood_2017}. 
	\looseness=-1
    
    	\begin{table}
		\caption{}\label{tab:blurmodels}
		\begin{tabu}{l | X | l}
			   & $\mathbf{B}_s$ & $\mathbf{B}_d$ \\
			\hline
			GPL-I & Identity (no blur) & Identity \\
			GPL-SI & Blur kernel from isocenter impulse response & Equation \eqref{eq:detmtf}\\
			GPL-SV & Different blur kernels for each projection, calculated from impulse at phantom location & Equation \eqref{eq:detmtf}\\
		\end{tabu}
	\end{table}
    
    Each reconstruction approach is a special case solution of (\ref{eq:objective}) using different approximations of system blur.
	The blur models used for each \ac{GPL} method are summarized in Table~\ref{tab:blurmodels}.
	\ac{GPL-I} did not include blur and is similar to a traditional \ac{MBIR} approach presuming uncorrelated Gaussian noise.
	\ac{GPL-SI} and \ac{GPL-SV} model scintillator blur (matched with that used in the data generation step) and focal spot blur kernels based on simulated impulse responses at particular locations.
	The impulse responses were generated using sourcelets as in the data generation step, but with sourcelet sampling reduced by a factor of two in each dimension.
	\ac{GPL-SI}'s focal spot blur model used a single blur kernel calculated from an impulse placed at isocenter (origin in Fig.~\ref{fig:fdk}A).
	\ac{GPL-SV} used a different blur kernel for each projection.
	Specifically, the location of the impulse response corresponded to the center of the phantom probe location.
	The blur kernel changes for every projection due to 1) change in magnification at the impulse location as a function of angle and 2) change in apparent focal spot shape due to varying obliquity of the ray from the source to the impulse.
	The focal spot blur model for each projection is shift-invariant, but because a different blur model (impulse response) was used for each projection, the overall effect of this blur model is shift-variant when reconstruction is applied.
	\looseness=-1
	
    All of these assumptions are a mismatch with the source model in data generation to varying degrees.
	However, the \ac{GPL-SV} model is an excellent approximation for a small region-of-interest since location-dependent effects are marginal and the variable magnification and angular-dependence of the apparent focal spot are handled by the view-dependent blur.
	Thus, the \ac{GPL-SV} investigation is a good indicator of regionally optimized performance, and potentially a good approximation of performance should a more sophisticated reconstruction with a global model for shift-variant blur be adopted (e.g., a full sourcelets model).
	\looseness=-1

	Following \cite{tilley:16:msv, tilley_ii_penalized-likelihood_2017}, $\mathbf{K}_Y^{-1}$ was applied using an iterative solution for \ac{GPL-SI} and \ac{GPL-SV}.
	Similarly, readout noise was assumed negligible in certain computation steps to avoid application of $\mathbf{K}_Y^{-1}$ every iteration, as described in \cite{tilley_nonlinear_2016, tilley_ii_penalized-likelihood_2017}.
	The term $\mathbf{K}_Y^{-1} \mathbf{y}$ was precomputed using \num{\Kniters{}} iterations of the preconditioned conjugate gradient method.
	All \ac{GPL} methods used a Huber Penalty \cite{huber_robust_statistics} with $\delta = \num{\huberdelta{}}$ and were optimized in terms of \ac{RMSE} over a range of penalty strengths ($\beta$'s).
	Each reconstruction ran for \niters{} iterations using \nsubsets{} subsets and Nesterov acceleration \cite{Nesterov2005, Kim2015}.
	The phantom was reconstructed using the three \ac{GPL} methods at the locations indicated by circles in Fig.~\ref{fig:fdk}A (\SI{80}{\milli\meter} from isocenter; \ang{0}, \ang{90}, \ang{180}, and \ang{360} about the $y$ axis; \cs{} and \as{}; and using both tube orientations).
	\looseness=-1

\section{Results}

\subsection{FDK Short Scan Maps}

The results of the FDK sweep can be seen in Fig.~\ref{fig:fdk}.
Fig.~\ref{fig:fdk}A is a schematic of the locations sampled, with each circular path corresponding to a constant radius in one of the plots in Fig.~\ref{fig:fdk}B-G.
The $y = \pm $ \SI{60}{\milli\meter} planes (Fig.~\ref{fig:fdk}B-C,F-G) have a substantially higher \ac{RMSE} than the \cs{} plane (Fig.~\ref{fig:fdk}D-E).
Because this \ac{RMSE} increase occurs with both X-ray tube orientations and is similar in magnitude above and below \cs{}, most of this image quality decrease is likely due to the large cone angle (e.g., incomplete sampling and cone-beam artifacts).
\looseness=-1

\begin{figure}[t!]
	\centering
	\input{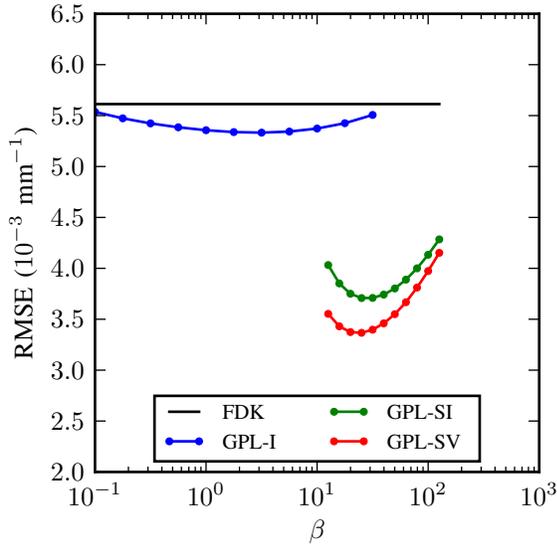}
	\caption{Example $\beta$ sweep results at the red dot in Fig.~\ref{fig:fdk}A with a transaxial focal spot orientation.}
	\label{fig:betasweep}
\end{figure}

Within the \cs{} plane, the transaxial orientation (Fig.~\ref{fig:fdk}D) exhibits more variance in image quality as compared to the axial orientation (Fig.~\ref{fig:fdk}E).
This is mainly a spatial resolution effect since the focal spot blur shift-variance is predominately along the anode-cathode axis,
which is in-plane in D (parallel to the $x$ axis at middle of the scan) but perpendicular to the plane in E (parallel to the $y$ axis).
For the transaxial orientation, the \ac{RMSE} varies more along the $x$ axis than the $z$ axis, as expected, with the lowest \ac{RMSE} on the left side of the image, corresponding to the side with a smaller apparent focal spot blur (i.e., the anode side).
In other words, as the source travels around isocenter, the apparent focal spot at the negative $x$ axis is generally smaller than at isocenter.
The exception to this being when the source is at the extremes of the scan, specifically, in the negative $z$ half plane.
However, this is a minority of the projection angles, and the net result is a smaller blur at an object on the $-x$ axis than at isocenter.
The axial orientation (Fig.~\ref{fig:fdk}E) shows little variation along the $x$ axis.
Fig.~\ref{fig:fdk}D-E also show depth-dependence effects, with the minimum \ac{RMSE} offset in the positive $z$ direction.
As the source orbits isocenter, the apparent focal spot blur size will change as magnification changes.
However, even though increasing magnification increases focal spot blur, it also decreases the effect of detector blur (scintillator blur and pixel sampling).
Thus there is an optimum magnification where these two effects are balanced. 
Placing the phantom at locations with an average magnification close to this optimum should result in superior image quality.
In this system, this optimum magnification appears to be more than the magnification at isocenter.
Thus, with a transaxial tube orientation, the best image quality is achieved by placing the phantom as far to the anode side of the \ac{FOV} as possible, and potential offset in $z$, while the optimum position with an axial orientation is along the $z$ axis, with the optimal location along the axis dependent on the tradeoff between focal spot blur, scintillator blur, and pixel sampling.
\looseness=-1

With the transaxial orientation, the $y = \pm $ \SI{60}{\milli\meter} planes exhibit a similar pattern, favoring the $-x$ direction to reduce focal spot blur.
The axial orientation shows a dramatic difference between these two planes, due to the anode-cathode axis being parallel to the $y$ axis. 
The overall \ac{RMSE} is lower in Fig.~\ref{fig:fdk}G (\SIr{5}{\lowv{}}{\per\milli\meter}) than Fig.~\ref{fig:fdk}C (\SIr{5}{\highv{}}{\per\milli\meter}), consistent with the fact that the apparent focal spot is smaller below the \cs{} plane.
The \ac{RMSE} values in Fig.~\ref{fig:fdk}G also vary dramatically as compared to Fig.~\ref{fig:fdk}C, as apparent focal spot size is more sensitive to magnification ($z$ position) on the anode side of the anode-cathode axis (i.e., $y < 0$) as compared to the cathode side.
\looseness=-1

With both orientations, the optimum location is partially dependent on the optimum magnification, which is a function of focal spot blur, scintillator blur, and pixel sampling.
With a transaxial tube orientation, position along the anode-cathode axis is particularly important, resulting in large in-plane variance in \ac{RMSE}.
Moving out of the central plane reduces image quality due to cone-beam artifacts and incomplete sampling.
With an axial orientation, in-plane variance is reduced.
However, image quality at different locations above and below the central plane is effected by shift-variant blur along the anode-cathode axis as well as cone beam effects.
Thus, the optimum position may be slightly off the \cs{} plane (in this case in the negative $y$ direction) because of the reduction in focal spot blur.
\looseness=-1

\subsection{MBIR}

Fig.~\ref{fig:betasweep} shows the results of the $\beta$ sweep in the \cs{} plane with a transaxial orientation. 
Comparing the minimum \ac{RMSE} for each reconstruction method, \ac{GPL-SV} results in the best image quality, followed by \ac{GPL-SI}, \ac{GPL-I}, and finally FDK.
The \ac{GPL-SI} and \ac{GPL-SV} methods show an increased sensitivity to $\beta$, likely due to the (noise amplifying) deblurring action of these methods.
\looseness=-1

\begin{figure}[t!]
	\centering
	\input{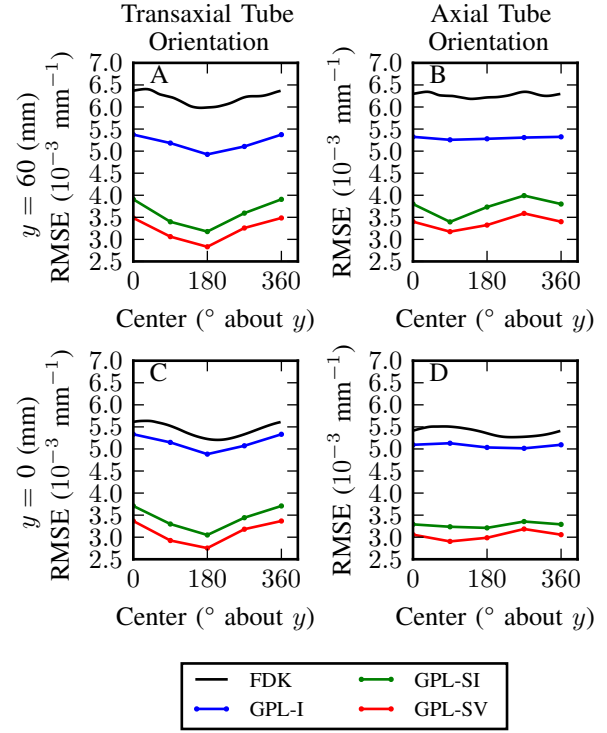}
	\caption{Minimum \ac{RMSE} from each $\beta$ sweep along with the FDK \ac{RMSE} values. Data is wrapped around at \ang{360} (i.e., the \ang{360} data are the same as the \ang{0} data).}
	\label{fig:rmse}
\end{figure}

\begin{figure*}
	\centering
	\input{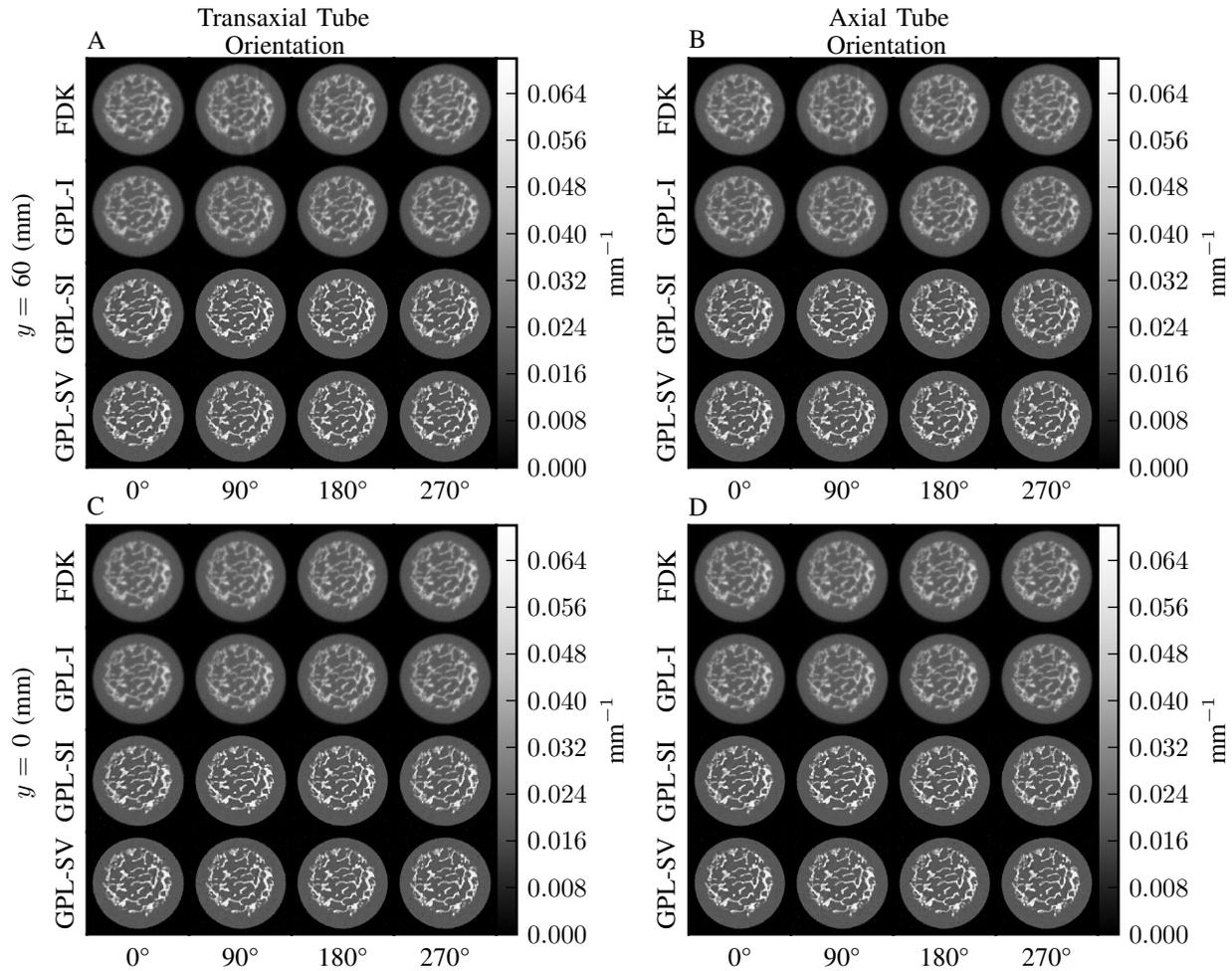}
	\caption{Axial slices of reconstructions. Each set of 16 reconstruction corresponds to a X-ray tube orientation (columns) and plane in the \ac{FOV} (rows). Within each set, columns correspond to the location of the phantom (in degrees rotated about the $y$ axis) and rows to reconstruction method.}
	\label{fig:axialrec}
\end{figure*}

\begin{figure*}
	\centering
	\input{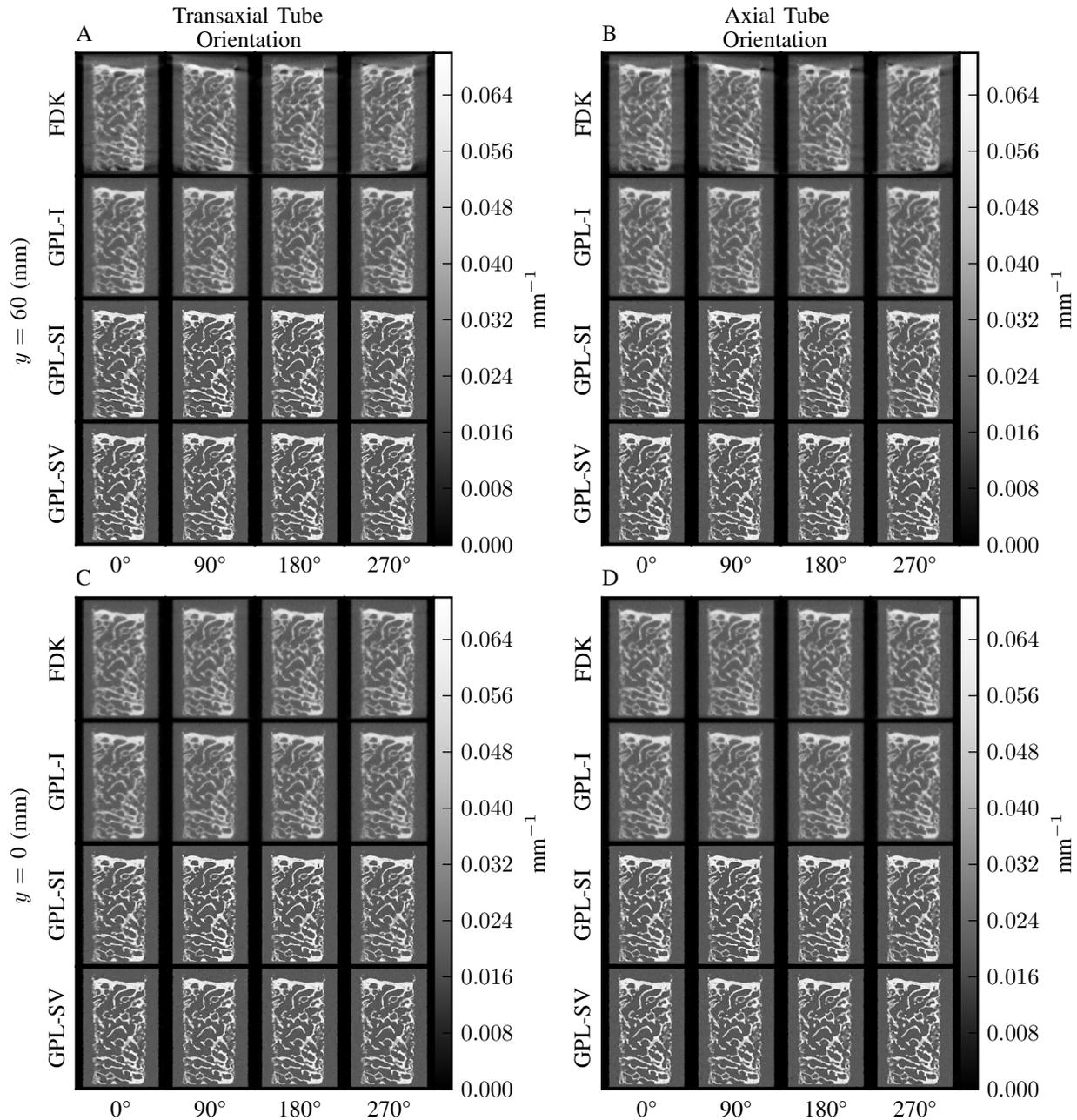}
	\caption{Coronal slices of reconstructions. Images are arranged as in Fig.~\ref{fig:axialrec}.}
	\label{fig:coronalrec}
\end{figure*}

We conducted $\beta$ sweeps similar to Fig.~\ref{fig:betasweep} for both X-ray tube orientations at \SI{80}{\milli\meter} from the axis of rotation in the $y =$ \SIlist{0;60}{\milli\meter} planes and at \ang{0}, \ang{90}, \ang{180}, and \ang{270} rotations (i.e., the green/red circles in Fig.~\ref{fig:fdk}A).
The minimum \ac{RMSE} across $\beta$'s was plotted as a function of angle (Fig.~\ref{fig:rmse}), along with the FDK reconstruction \acp{RMSE}.
In all cases, the rank ordering of performance across reconstruction methods is the same as in Fig.~\ref{fig:betasweep}, with \ac{GPL-SV} providing the best image quality.
The most dramatic improvement is between \ac{GPL-I} and \ac{GPL-SI}, due to the introduction of the blur model.
Improving the accuracy of the blur model with \ac{GPL-SV} further improves image quality as compared to \ac{GPL-SI}.
\looseness=-1

Unlike FDK, the \ac{GPL} methods result in similar \ac{RMSE} values in the \cs{} plane (Fig.~\ref{fig:rmse}C,D) and the \as{} plane (Fig.~\ref{fig:rmse}A,B).
This is likely due to the adavantages of the Huber regularizer which better handles the incomplete data at high cone angles.

With the axial orientation and in the \as{} plane (Fig.~\ref{fig:rmse}B), the blur at the axis of rotation is larger than at isocenter.
Therefore at \ang{0} and \ang{180} \ac{GPL-SI} generally underestimates the blur, causing \ac{GPL-SI} to have reduced image quality as compared to \ac{GPL-SV}.
At \ang{90}, the blur is reduced relative to the blur at \ang{0} and \ang{180} due to a decrease in magnification, decreasing \ac{RMSE} in both \ac{GPL-SI} and \ac{GPL-SV}.
Additionally, the \ac{GPL-SI} blur model is more accurate at this location, resulting in the similar performance of \ac{GPL-SI} and \ac{GPL-SV}.
At \ang{270} the reverse occurs, and the \ac{RMSE} values of both methods increase.
While \ac{GPL-SI} further underestimates the blur, in this experiment there was not a substantial increase in the \ac{RMSE} difference between \ac{GPL-SI} and \ac{GPL-SV} at \ang{270} as compared to \ang{0} and \ang{180}.
\ac{GPL-I} and FDK both exhibited less variance in \ac{RMSE} with angle than \ac{GPL-SI} and \ac{GPL-SV}.
\looseness=-1

The \ac{RMSE} results are supported by a qualitative analysis of the reconstructions (Fig.~\ref{fig:axialrec} and Fig.~\ref{fig:coronalrec}).
The reconstructions show the dramatic improvement due to incorporation of the blur model.
In all cases (both tube orientations and both planes), the FDK and \ac{GPL-I} reconstructions (top two rows), have substantially more blur than the \ac{GPL-SI} and \ac{GPL-SV} reconstructions (bottom two rows).
The FDK coronal slices in the \as{} plane (Fig.~\ref{fig:coronalrec}A-B) have substantial cone-beam artifacts near the top of the phantom, as compared to the \ac{GPL} coronal slices.
In the \cs{} plane with the transaxial orientation (Fig.~\ref{fig:axialrec}C and Fig.~\ref{fig:coronalrec}C), the FDK reconstruction at \ang{180} is sharper than the reconstruction at \ang{0}, consistent with the \ac{RMSE} results.
Differences between \ac{GPL-SI} and \ac{GPL-SV} are subtle, but are apparent when comparing how the performance of these two methods vary with angle.
For example, with an axial tube orientation and in the \as{} plane (Fig.~\ref{fig:axialrec}B and Fig.~\ref{fig:coronalrec}B), \ac{GPL-SI} produces the sharpest reconstruction at \ang{90} (consistent with the \ac{RMSE} results).
On the other hand, the \ac{GPL-SV} method produces visually similar reconstructions.
This is particularly evident in the coronal slices (Fig.~\ref{fig:coronalrec}B).
\looseness=-1

In general, axial tube orientations result in more uniform image quality, while transaxial tube orientations result in better image quality in a specific region of interest, at the cost of reduced image quality elsewhere.
\ac{MBIR} and blur modeling were able to improve image quality as compared to FDK in all cases, with more accurate models resulting in more accurate reconstructions.
Even without blur modeling, \ac{GPL-I} was able to reduce cone-beam artifact and improve image quality at high cone angles as compared to FDK, resulting in more uniform image quality as a function of slice.
However, the best results were obtained with the accurate, shift-variant model of focal spot blur.
\looseness=-1

\section{Discussion}

By modeling an extremities \ac{CBCT} imaging system we have shown that shift-variant focal spot blur can cause image quality to vary with position throughout an \ac{FOV}.
Understanding these properties can aid design of application and system specific acquisition protocols.
For example, when imaging a small object on this system with a short scan and reconstructing using FDK, the best image quality will be obtained with a transaxial X-ray tube orientation and the object at the edge of the \ac{FOV} one the anode side of the anode--cathode axis ($-x$ in \ref{fig:fdk}).
Similarly, placing the object at the cathode side should be avoided.
On the other hand, when scanning a large object which requires image-quality uniformity, an axial X-ray tube orientation should be used.
The results show that \ac{MBIR} methods may alter these trade-offs depending on the incorporated blur models.
For example, with an axial X-ray tube orientation and an object in the \as{} plane, position of the object is more important when using \ac{GPL-SI} as compared to \ac{GPL-I} or FDK when minimizing \ac{RMSE}.
In all cases the more accurate the blur model used in the reconstruction algorithm, the more accurate the reconstruction.
While we only analyzed a single \ac{CBCT} system, the methods used may be applied to a wide of range geometries, focal spots, target resolutions, etc.
\looseness=-1

The \ac{GPL-SV} method assumes blurs are shift-invariant for each projection.
Without this assumption, a full sourcelets model is required (i.e., modeling each sourcelet in the \ac{MBIR} algorithm).
The \ac{GPL} method can incorporate such a model by increasing the number of rows in $\mathbf{A}$ by a factor of the number of sourcelets, and adding a weighted sum operation to the $\mathbf{B}$ matrix.
This increases computation time by (approximately) a factor of the number of sourcelets.
\ac{GPL-SV} is substantially faster and, for small phantoms, roughly equivalent to the full sourcelets model.
Thus, \ac{GPL-SV} cheaply assesses how well a full sourcelets \ac{MBIR} algorithm will perform.
In future work we will compare \ac{GPL-SV} to a full sourcelets model to confirm these assumptions and to reconstruct large objects.
The \ac{GPL-SV} method may also be used when high resolution is only required in a small \ac{ROI}, in which case a shift-variant model may be used for the \ac{ROI} and a simpler model may be used for the remainder of the object \cite{cao_multiresolution_2016}.
\looseness=-1

The optimal acquisition protocol and reconstruction method is dependent on the imaging task as well as system properties.
Therefore, task specific image quality metrics may provide more application specific information.
This work used \ac{RMSE} because it is general and well understood, but future work will consider other metrics.
Reconstructions in this work exclusively used short scan data, allowing angle dependent image quality to be analyzed (as compared to full scan reconstructions, in which all locations at constant $y$ and radius are equivalent).
However, full scans with \ac{GPL-SV} would allow data from high and low resolution portions of the scan path to be combined efficiently, as suggested in previous work \cite{tilley:16:msv}.
We will reassess full scans with shift-variant modeling in future work.
\looseness=-1

Through analysis of shift-variant blur properties of a \ac{CBCT} system,
we have demonstrated how understanding location-dependent blur can improve acquisition protocol design
(e.g., favoring the anode side of the X-ray source)
and system design (e.g., axial tube orientation for uniformity, transaxial for \ac{ROI} imaging).
It was also shown that proper modeling of shift-variant blur properties with \ac{MBIR} leads to improved image quality over traditional methods.
As CT systems target increasingly high-resolution applications, understanding and accounting for shift-variant system blurs will be essential to providing better accuracy and improved clinical outcomes.
\looseness=-1

\section*{Acknowledgments}

This work was partly funded by United States National Institutes of Health grants R01~EB018896 and F31~EB023783.
The authors would like to express their gratitude to Yoshi Otake and Ali \"{U}neri for developing the GPU routines used in this work.
Most reconstructions were done on the Maryland Advanced Research Computing Center cluster.

\bibliography{ZoteroLibrarybibtex}

\begin{thebibliography}{10}
\providecommand{\url}[1]{#1}
\csname url@samestyle\endcsname
\providecommand{\newblock}{\relax}
\providecommand{\bibinfo}[2]{#2}
\providecommand{\BIBentrySTDinterwordspacing}{\spaceskip=0pt\relax}
\providecommand{\BIBentryALTinterwordstretchfactor}{4}
\providecommand{\BIBentryALTinterwordspacing}{\spaceskip=\fontdimen2\font plus
\BIBentryALTinterwordstretchfactor\fontdimen3\font minus
  \fontdimen4\font\relax}
\providecommand{\BIBforeignlanguage}[2]{{%
\expandafter\ifx\csname l@#1\endcsname\relax
\typeout{** WARNING: IEEEtran.bst: No hyphenation pattern has been}%
\typeout{** loaded for the language `#1'. Using the pattern for}%
\typeout{** the default language instead.}%
\else
\language=\csname l@#1\endcsname
\fi
#2}}
\providecommand{\BIBdecl}{\relax}
\BIBdecl

\bibitem{Lai2007}
C.-J. Lai, C.~C. Shaw, L.~Chen, M.~C. Altunbas, X.~Liu, T.~Han, T.~Wang, W.~T.
  Yang, G.~J. Whitman, and S.-J. Tu, ``Visibility of microcalcification in cone
  beam breast {{CT}}: Effects of {{X}}-ray tube voltage and radiation dose.''
  \emph{Medical Physics}, vol.~34, no.~7, pp. 2995--3004, 2007.

\bibitem{Kwan2007}
A.~L.~C. Kwan, J.~M. Boone, K.~Yang, and S.-Y. Huang, ``Evaluation of the
  spatial resolution characteristics of a cone-beam breast {{CT}} scanner.''
  \emph{Medical Physics}, vol.~34, no.~1, pp. 275--281, 2007.

\bibitem{Carrino2014}
J.~A. Carrino, A.~Al~Muhit, W.~Zbijewski, G.~K. Thawait, J.~W. Stayman,
  N.~Packard, R.~Senn, D.~Yang, D.~H. Foos, J.~Yorkston, and J.~H. Siewerdsen,
  ``Dedicated cone-beam {{CT}} system for extremity imaging.''
  \emph{Radiology}, vol. 270, no.~3, pp. 816--24, 2014.

\bibitem{marinetto_quantification_2016}
E.~Marinetto, M.~Brehler, A.~Sisniega, Q.~Cao, J.~W. Stayman, J.~Yorkston,
  J.~H. Siewerdsen, and W.~Zbijewski, ``Quantification of bone
  microarchitecture in ultrahigh resolution extremities conebeam {{CT}} with a
  {{CMOS}} detector and compensation of patient motion,'' in \emph{Computer
  {{Assisted Radiology}} 30th {{International Congress}} and {{Exhibition}}},
  Heidelberg, Germany, jun 2016.

\bibitem{xu_technical_2016}
J.~Xu, A.~Sisniega, W.~Zbijewski, H.~Dang, J.~W. Stayman, M.~Mow, X.~Wang,
  D.~H. Foos, V.~E. Koliatsos, N.~Aygun, and J.~H. Siewerdsen,
  ``\BIBforeignlanguage{en}{Technical assessment of a prototype cone-beam
  {{CT}} system for imaging of acute intracranial hemorrhage},''
  \emph{\BIBforeignlanguage{en}{Medical Physics}}, vol.~43, no.~10, pp.
  5745--5757, oct 2016.

\bibitem{Tsui1987}
B.~M.~W. Tsui, H.~B. Hu, D.~R. Gilland, and G.~T. Gullberg, ``Implementation of
  {{Simultaneous Attenuation}} and {{Detector Response Correction}} in
  {{Spect}}.'' \emph{IEEE Transactions on Nuclear Science}, vol.~35, no.~1, pp.
  778--783, 1987.

\bibitem{mumcuoglu_accurate_1996}
E.~U. Mumcuoglu, R.~M. Leahy, S.~R. Cherry, and E.~Hoffman, ``Accurate
  geometric and physical response modelling for statistical image
  reconstruction in high resolution {{PET}},'' in \emph{, 1996 {{IEEE Nuclear
  Science Symposium}}, 1996. {{Conference Record}}}, vol.~3, nov 1996, pp.
  1569--1573 vol.3.

\bibitem{LaRiviere2007}
P.~J. La~Rivi{\`e}re and P.~Vargas, ``Correction for resolution nonuniformities
  caused by anode angulation in computed tomography,'' \emph{IEEE Transactions
  on Medical Imaging}, vol.~27, no.~9, pp. 1333--1341, 2008.

\bibitem{tilley:16:msv}
S.~Tilley~II, W.~Zbijewski, J.~H. Siewerdsen, and J.~W. Stayman, ``Modeling
  shift-variant {{X}}-ray focal spot blur for high-resolution flat-panel
  cone-beam {{CT}},'' in \emph{Proc. 4th {{Intl}}. {{Mtg}}. on Image Formation
  in {{X}}-Ray {{CT}}}, 2016.

\bibitem{tilley_ii_penalized-likelihood_2017}
S.~Tilley~II, M.~Jacobson, Q.~Cao, M.~Brehler, A.~Sisniega, W.~Zbijewski, and
  J.~W. Stayman, ``Penalized-{{Likelihood Reconstruction}} with
  {{High}}-{{Fidelity Measurement Models}} for {{High}}-{{Resolution
  Cone}}-{{Beam Imaging}},'' \emph{IEEE Transactions on Medical Imaging
  (submitted)}, 2017.

\bibitem{tilley_nonlinear_2016}
S.~Tilley~II, J.~H. Siewerdsen, W.~Zbijewski, and J.~W. Stayman, ``Nonlinear
  statistical reconstruction for flat-panel cone-beam {{CT}} with blur and
  correlated noise models,'' in \emph{{{SPIE}} 9783 {{Medical Imaging}} 2016:
  {{Physics}} of {{Medical Imaging}}}, vol. 9783, 2016, pp.
  97\,830R--97\,830R--6.

\bibitem{huber_robust_statistics}
P.~J. Huber, \emph{Robust Statistics}.\hskip 1em plus 0.5em minus 0.4em\relax
  New York: Wiley, 1981.

\bibitem{Nesterov2005}
Y.~Nesterov, ``Smooth minimization of non-smooth functions,''
  \emph{Mathematical Programming Journal, Series A}, vol. 103, pp. 127--152,
  2005.

\bibitem{Kim2015}
D.~Kim, S.~Ramani, and J.~A. Fessler, ``Combining {{Ordered Subsets}} and
  {{Momentum}} for {{Accelerated X}}-{{Ray CT Image Reconstruction}},''
  \emph{IEEE Transactions on Medical Imaging}, vol.~34, no.~1, pp. 167--178,
  2015.

\bibitem{cao_multiresolution_2016}
Q.~Cao, W.~Zbijewski, A.~Sisniega, J.~Yorkston, J.~H. Siewerdsen, and J.~W.
  Stayman, ``\BIBforeignlanguage{en}{Multiresolution iterative reconstruction
  in high-resolution extremity cone-beam {{CT}}},''
  \emph{\BIBforeignlanguage{en}{Physics in Medicine and Biology}}, vol.~61,
  no.~20, p. 7263, 2016.

\end{thebibliography}

\end{document}